\documentclass[twocolumn,amsmath,amssymb,nofootinbib,a4paper]{revtex4} 

\RequirePackage{graphicx}
\usepackage{amsmath}
\usepackage{xcolor}
\usepackage[normalem]{ulem}
\usepackage{units}

\newcommand{\pt}{p_\perp}

\begin{document}

\graphicspath{{Images}}

\title{Effects of Multi-Parton Interactions in Jet Quenching in Heavy-Ion Collisions}

\author{Andrecia Ramnath}
\email{andrecia.ramnath@fysik.lu.se}
\affiliation{Department of Physics, Lund University, Box 118, SE 22100 Lund, Sweden }

\author{Korinna Zapp}
\email{korinna.zapp@fysik.lu.se}
\affiliation{Department of Physics, Lund University, Box 118, SE 22100 Lund, Sweden }

\begin{abstract}
We perform the first systematic study of the effects of multi-parton interactions (MPI's) in the context of jet quenching in heavy-ion collisions with the jet quenching model \textsc{Jewel}. We use the simple MPI model of \textsc{Pythia}\,6, on which \textsc{Jewel} is based. We find negligible effects on all observables except jet--hadron and $Z$--hadron correlations, which show a moderate enhancement at large distances. More detailed analysis at parton level reveals that, in heavy-ion collisions, the MPI contribution to jets is suppressed by quenching effects.
\end{abstract}

\maketitle

\section{Introduction}
\label{sec::intro}

The substructure of quenched jets is the subject of intense research both theoretically~\cite{Andrews:2018jcm,Cao:2020wlm} and experimentally~\cite{Armesto:2015ioy,Connors:2017ptx,CMS:2024krd}. Medium-induced radiation~\cite{Blaizot:2014ula,Chien:2015hda,Casalderrey-Solana:2015bww,Dominguez:2019ges,Arnold:2020uzm,Mehtar-Tani:2021fud}, color coherence~\cite{Casalderrey-Solana:2011ule,Mehtar-Tani:2011hma,Barata:2021byj,Andres:2022ndd} and medium response~\cite{Neufeld:2011yh,Tachibana:2014lja,He:2015pra,Tachibana:2015qxa,Wang:2013cia,Cao:2016gvr,Casalderrey-Solana:2016jvj,KunnawalkamElayavalli:2017hxo,Milhano:2017nzm} are expected to leave imprints on the internal structure of jets. The ultimate goal is thus to decode this information about the microscopic workings of parton--medium interactions. However, to understand the sub-structure of quenched jets is challenging both from a theoretical and an experimental perspective. In an attempt to get possibly confounding factors under control, the effect of initial-state radiation was studied in~\cite{Zapp:2022dhq}. Here, continue that effort by investigating the effect of multi-parton interactions (MPI's) in the context of jet quenching.

MPI's arise when, at high centre-of-mass energies, the probability of having more than one parton--parton scattering in a proton--(anti-)proton collision becomes sizable. Formally, this effect is beyond standard factorization theorems and therefore phenomenological modeling is needed. In Monte Carlo event generators, MPI's are simulated as secondary $2\to2$ partonic scatterings in QCD, as they are expected to be perturbatively hard. MPI's give rise to semi-hard hadronic activity that is largely uncorrelated with the hard scattering and is observed in the form of the underlying event.

MPI's have been studied extensively in proton--proton collisions~\cite{Bartalini:2018qje}, but have not been a main focus in heavy-ion physics. An exception is~\cite{Yang:2021qtl}, where it was observed that MPI activity on the $Z$ boson side of $Z$+jet events can obscure signs of the diffusion wake. 

\section{Modeling jet evolution and the underlying event in JEWEL}
\label{sec::model}

\textsc{Jewel}~\cite{Zapp:2012ak,Zapp:2013vla} relies heavily on \textsc{Pythia}\,6.4~\cite{Sjostrand:2006za} for the event generation. In particular, the hard scattering matrix elements, initial state parton showers including PDF handling and hadronization are provided by \textsc{Pythia}\,6.4. We therefore let \textsc{Pythia}\,6.4 also generate the additional MPI  scatterings. There are two versions of the MPI model in \textsc{Pythia}\,6.4: the so-called `old model'~\cite{Sjostrand:1987su} and the `new model'~\cite{Sjostrand:2004ef} in which generation of the $\pt$ ordered sequence of MPI scatterings is interleaved with the parton shower evolution. Since \textsc{Jewel} has its own parton shower, the interleaved model does not work with \textsc{Jewel} and we instead use the `old model'. While this is a simpler model than the more sophisticated one developed later, it can still inform us whether sizable effects from MPI's can be expected in the context of jet quenching studies. Here, we will summarize the main features of the model; for a more detailed discussion the reader is referred to~\cite{Sjostrand:1987su,Sjostrand:2006za}.

The jet cross section above some minimal transverse momentum $\pt^\text{min}$ is given by
\begin{equation}
\sigma_\mathrm{hard} = \int \limits_{(\pt^\text{min})^2}^{s/4} \frac{\mathrm{d}\,\sigma}{\mathrm{d}\,\pt^2} \,\mathrm{d}\,\pt^2 \,. 
\end{equation}
This cross section diverges for $\pt^\text{min} \to 0$ and saturates the non-diffractive proton--proton cross section $\sigma_\mathrm{nd}$ for perurbatively high values of $\pt^\text{min}$ at sufficiently high collider energies. Since $\sigma_\mathrm{hard}$ is a partonic cross section, this is interpreted as a sign for several partonic scatterings taking place in one proton--proton collision. These are postulated to be independent of each other, so that the number of parton--parton scatterings follows a Poisson distribution with mean 
\begin{equation}
    \bar n = \frac{\sigma_\mathrm{hard}}{\sigma_\mathrm{nd}} \,. 
    \label{eq:meannmpi}
\end{equation}
The scatterings are generated as a sequence with falling $\pt$. In each step, the PDF's are rescaled to take into account the energy already taken out by the previous scatterings. This way the hardest scattering is guaranteed to be unmodified by subsequent scatterings. The color treatment of the MPI scatterings is simplified in order to avoid too complicated color topologies.

There are different options for modeling the matter distribution inside the proton, the default being a double Gaussian with a narrow core representing the valence quarks surrounded by a broader distribution of gluons and sea quarks. The mean number of MPI scatterings is then dependent on the impact parameter, i.e.\ the transverse distance between the cores of the colliding protons. In practice, the mean number of scatterings at a given impact parameter is taken to be proportional to the matter overlap in such a way that, averaging over impact parameters, the original relation eq.~\ref{eq:meannmpi} is recovered.

In the \textsc{Pythia} implementation of the model, MPI scatterings have no parton showers. When running with \textsc{Jewel}, there is an option to supplement the MPI scatterings with final-state parton showers. The partons from MPI scatterings are showered pairwise such that the recoil is transferred only between partons coming from the same scattering. For the starting scale for the parton shower, we use~\cite{Ellis:1992en}
\begin{equation}
    Q_\mathrm{max} = \frac{\pt e^{0.3 \Delta y/2}}{2} \,,
\end{equation}
where $\Delta y$ is the rapidity difference between the two outgoing partons of an MPI scattering (for the hardest scattering the starting scale is still just the $\pt$ of the hard scattering).

The (semi-)hard partons produced by secondary MPI scatterings interact in the dense background in the same way as the partons coming from the hardest scattering. In \textsc{Jewel} the small differences in the production points of the partons coming from different MPI scatterings are neglected, i.e. all partons coming out of hard scatterings are placed at the same production point. The partons then propagate in the background medium and undergo elastic scattering. If a scattering is hard enough it should have radiative corrections generated by a parton shower. In practice, a trial parton shower with starting scale given by the hardness of the scattering is started. If the first emission of the trial parton shower has a formation time that is shorter than that of the next emission from the original parton shower, the trial parton shower becomes the new parton shower and the old one is stopped. Otherwise, the trial parton shower is rejected and the old one continues. Scattering in the background medium can thus enlarge the phase space for radiation, which leads to more radiation than in vacuum. When several scatterings occur during the formation time of an emission, they are added coherently in a way that reproduces the non-Abelian Landau--Pomerantchuk--Migdal (LPM) effect~\cite{Zapp:2008af}. After all scattering and radiation processes have terminated, the event is hadronized using the Lund string fragmentation model implemented in \textsc{Pythia}.

In \textsc{Jewel}, hard partons scatter off quasi-particles in the medium. There is no complete simulation of the background. Instead, a background parton is generated when one is needed for a scattering with a hard parton. This background parton then takes the recoil from the scattering. Since background partons are typically softer than the hard partons, the result of a scattering is usually that the hard parton loses energy and the background parton gains energy. There is an option in \textsc{Jewel} to keep the recoiling background parton in the event to get an estimate of where the energy lost by hard partons is going~\cite{KunnawalkamElayavalli:2017hxo}. This is only approximate, since the recoiling background partons do not interact in the medium themselves but simply free-stream. When the recoil partons are included in the event, their momentum prior to the scattering has to be removed from the final jets because uncorrelated background is subtracted from the jets. We here use the constituent subtraction scheme for removal of thermal momenta~\cite{Milhano:2022kzx}.

As far as we are aware, this is the first systematic study of the effect of MPI's in jet quenching.

\section{Results}
\label{sec::results}

$Z$+jet and di-jet samples of 500,000 each are generated for Pb+Pb and p+p collisions at $\sqrt{s_{NN}} = \unit[5.02]{TeV}$ using PDF sets provided by \textsc{Lhapdf\,6} \cite{Buckley:2014ana}.  Hadron-level results are generated using the \textsc{Epps16nlo} nuclear PDF set \cite{Eskola:2016oht} and parton-level results are generated with \textsc{Ct14nlo} PDF's \cite{Dulat:2015mca}. In this way, nuclear PDF's are not used to isolate the effects of MPI's. All events are analyzed with Rivet \cite{Bierlich:2019rhm} and the \textsc{FastJet} package~\cite{Cacciari:2011ma}.

In addition to the effect of MPI's, final-sate radiation (FSR) off the MPI's is also considered. In all plots, the red lines correspond to collisions with no MPI's. The blue lines correspond to collisions with MPI's but no final-state radiation off the MPI's. The green lines correspond to collisions with both MPI's and final-state radiation off the MPI's. In all results, medium response with constituent subtraction at event level, i.e.\ before jet clustering, is included.

For the background medium, we use \textsc{Jewel}'s standard simplified background model with initial temperature $T_\mathrm{i}=\unit[590]{MeV}$ at $\tau_\mathrm{i}=\unit[0.4]{fm}$~\cite{Shen:2014vra}.

%--------------------------------------------------

\subsection{$Z$+jet events}
\label{sec::results:z-jet}

Events in which charged hadrons produced in a parton shower from the same hard scattering as a leptonically decaying $Z$ boson are analysed. The results are compared to measurements by CMS \cite{CMS:2021otx}. The samples are generated with 0-30 \% centrality, where the largest modifications due to medium effects are expected. The analysis procedure matches as far as possible the experimental one.  Specifically, jets are reconstructed using the anti-$k_\perp$ algorithm~\cite{Cacciari:2008gp} and a jet radius of $R=0.4$ is chosen. $Z$ bosons with invariant mass $\unit[60]{GeV} < M_\text{Z} < \unit[120]{GeV}$ and transverse momentum $\pt^\text{Z} > \unit[30]{GeV}$ are considered. We only consider the decay of the $Z$ to muons, where in \cite{CMS:2021otx}, the $Z \rightarrow \mu^+\mu^-$ trigger has cutoffs of $\pt > \unit[12]{GeV}$ and $|\eta| < 2.4$ on one muon. 

\begin{figure}
\includegraphics[width=0.5\textwidth]{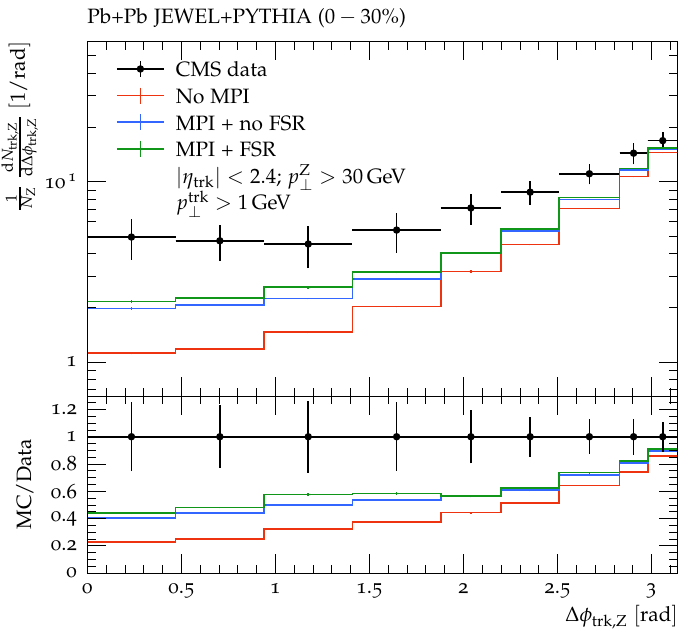}
\caption{Distribution of the angular separation $\Delta \phi_\text{trk,Z}$ between $Z$ bosons and charged-particle tracks in $Z$+jet events with $R=0.4$ in Pb+Pb collisions, as measured by CMS \cite{CMS:2021otx}. Only charged hadrons with $\pt > \unit[1]{GeV}$ are included.}
\label{fig:pbpb-deltaphi}
\end{figure}

\begin{figure}
\includegraphics[width=0.5\textwidth]{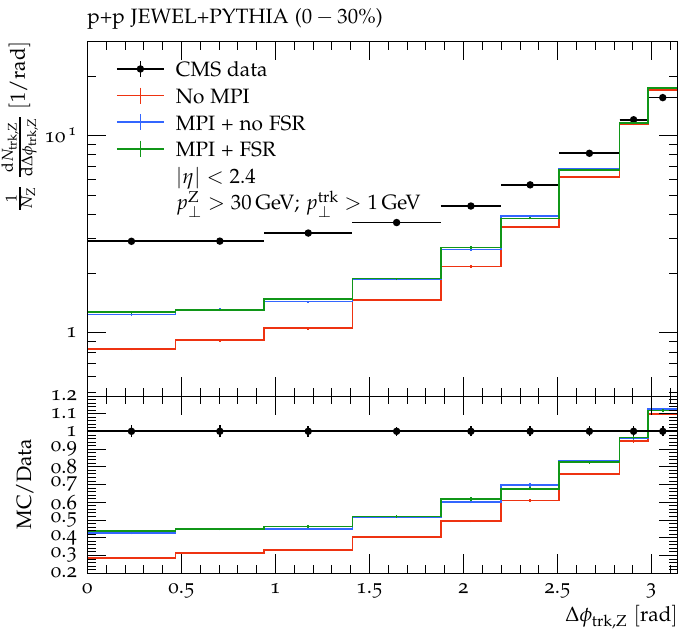}
\caption{Distribution of the angular separation $\Delta \phi_\text{trk,Z}$ between $Z$ bosons and charged-particle tracks in $Z$+jet events with $R=0.4$ in p+p collisions, as measured by CMS \cite{CMS:2021otx}. Only charged hadrons with $\pt > \unit[1]{GeV}$ are included.}
\label{fig:pp-deltaphi}
\end{figure}

Figures \ref{fig:pbpb-deltaphi} and \ref{fig:pp-deltaphi} show the distributions of the angular separation $\Delta \phi := |\phi_\text{trk} - \phi_Z|$ in Pb+Pb and p+p collisions, respectively. Here, $\phi_\text{Z}$ and $\phi_\text{trk}$ are the azimuthal angles of the $Z$ boson and of other charged tracks in the event, respectively. The distributions $\mathrm{d}N_\text{trk,Z}/\mathrm{d}\Delta\phi_\text{trk,Z}$ are normalised by $N_\text{Z}$, the number of Z bosons. There is an enhancement of the distribution on the boson side due to MPI's in both Pb+Pb and p+p collisions. This is in agreement with the findings in \cite{Yang:2021qtl}, where a coupled linear Boltzmann transport and hydro model is used to study the enhancement of soft hadrons in the direction of both the $Z$ boson and the jet. Figure \ref{fig:pbpb-deltaphi} shows that, in the Pb+Pb case, including FSR appears to enhance the spectrum even further. These findings raise the question whether MPI contributions are visible also in other jet observables that are sensitive to soft or semi-hard particles. We thus move on to examine a selection of such quantities.

\begin{figure}
\includegraphics[width=0.5\textwidth]{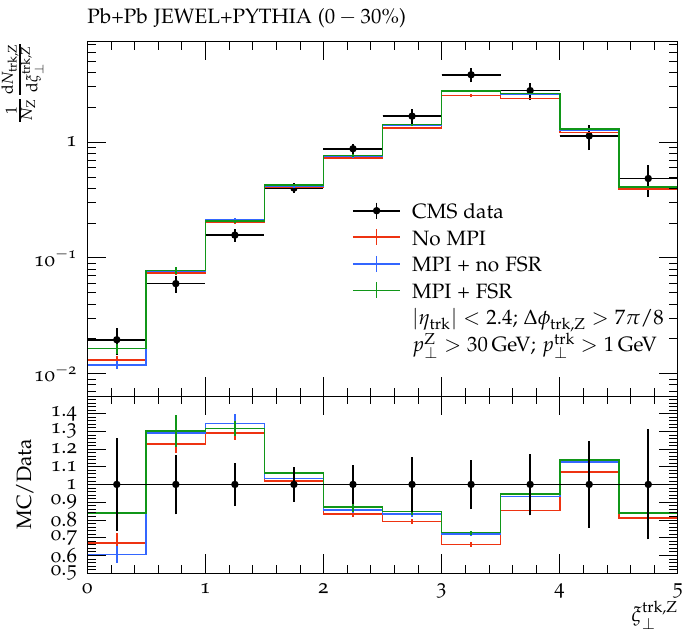}
\caption{Distribution of the jet fragmentation variable $\xi_\perp^\text{trk,Z}$ in $Z$+jet events with $R=0.4$ in Pb+Pb collisions, as measured by CMS \cite{CMS:2021otx}. Only charged hadrons with $\pt > \unit[1]{GeV}$ and $\Delta \phi_\text{trk,Z} > 7\pi/8$ are included.}
\label{fig:pbpb-xi}
\end{figure}

\begin{figure}
\includegraphics[width=0.5\textwidth]{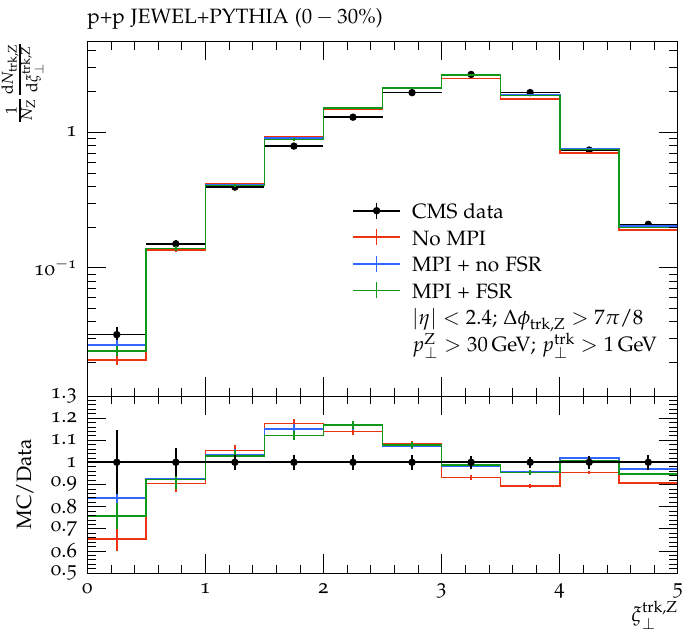}
\caption{Distribution of the jet fragmentation variable $\xi_\perp^\text{trk,Z}$ in $Z$+jet events with $R=0.4$ in p+p collisions, as measured by CMS \cite{CMS:2021otx}. Only charged hadrons with $\pt > \unit[1]{GeV}$ and $\Delta \phi_\text{trk,Z} > 7\pi/8$ are included.}
\label{fig:pp-xi}
\end{figure}

Figures \ref{fig:pbpb-xi} and \ref{fig:pp-xi} show the normalized distributions $1/N_\text{Z}\, \mathrm{d}N_\text{trk,Z}/\mathrm{d}\xi_\perp^\text{trk,Z}$ for the jet fragmentation variable $\xi_\perp^\text{trk,Z}$ in Pb+Pb and p+p collisions, respectively. This variable is the longitudinal momentum distribution of tracks on the jet side with $\Delta \phi_\text{trk,Z} > 7\pi/8$ and is defined as 
%----------
\begin{equation}
\xi_\perp^\text{trk,Z} :=
\ln \left( - \frac{|\vec{p}_\perp^\text{Z}|^2}{\vec{p}_\perp^{\,\text{trk}} \cdot \vec{p}_\perp^\text{Z}} \right).
\label{eq:xi}
\end{equation}
%----------
Here, $\vec{p}_\perp^\text{Z}$ and $\vec{p}_\perp^\text{trk}$ are the transverse momentum vectors (with respect to the beam direction) of the $Z$ boson and charge-particle track, respectively.

The results show a very slight increase at the high $\xi_\perp^\text{trk,Z}$ (low track $\pt$) end of the distribution due to MPI's. The effect is small since particles from MPI's are distributed uniformly in azimuthal angle relative to the $Z$. The region $\Delta \phi_\text{trk,Z} > 7\pi/8$ is dominated by the fragmentation of the jet and the MPI's contribution is relatively very small.
 
%--------------------------------------------------

\subsection{Di-jet hadron-level results}
\label{sec::results:hadrons}

The $0-10$ \% centrality interval is chosen for the di-jet samples. The hadron-level results shown here correspond to a jet radius of $R=0.4$. 

\begin{figure}
\includegraphics[width=0.5\textwidth]{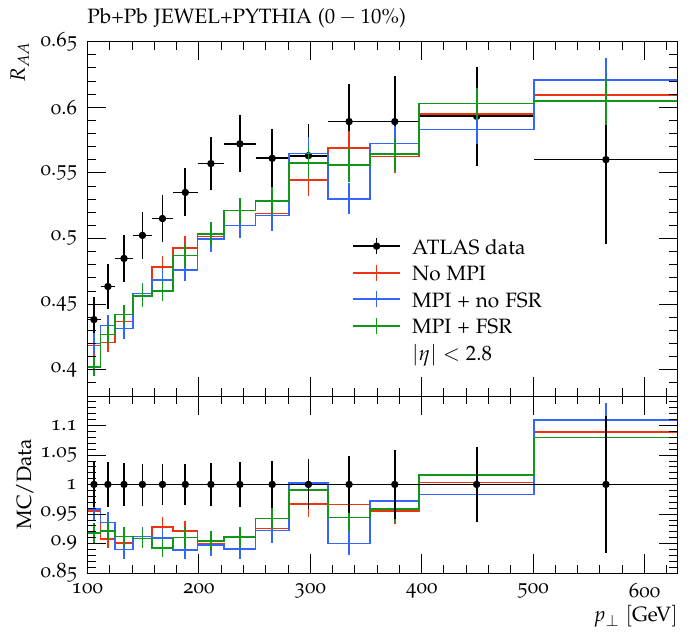}
\caption{Nuclear modification factor $R_{AA}$ for $|\eta|<2.8$ with $R=0.4$, as measured by \textsc{Atlas} \cite{ATLAS:2018gwx}.}
\label{fig:pbpb-raa}
\end{figure}

Figure \ref{fig:pbpb-raa} shows the nuclear modification factor $R_{AA}$ for $|\eta| < 2.8$ as measured by \textsc{Atlas} \cite{ATLAS:2018gwx}. No statistically significant modification due to MPI's is observed. MPI effects partially cancel in the ratio between p+p and Pb+Pb, and the jet $\pt$ is dominated by particles from the hardest scattering.

\begin{figure}
\includegraphics[width=0.5\textwidth]{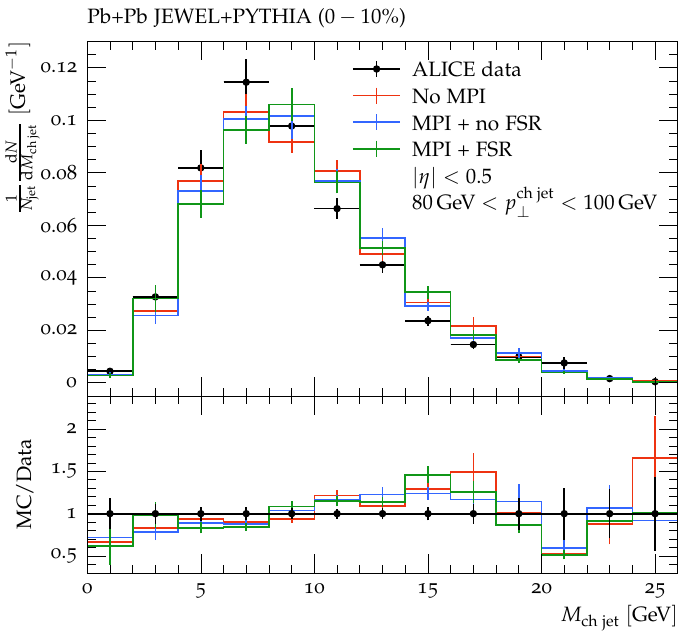}
\caption{Distribution for the charged-jet mass $M_\text{ch jet}$ with $R=0.4$ in Pb+Pb collisions, as measured by \textsc{Alice} \cite{ALICE:2017nij}. In  \cite{ALICE:2017nij}, reconstructed tracks are required to have $\pt^\text{trk} > \unit[0.15]{GeV}$}
\label{fig:pbpb-mass}
\end{figure}

An observable that is much more sensitive to soft particles at the periphery of the jet is the jet mass. Figure \ref{fig:pbpb-mass} shows the distribution for the charged-jet mass $M_\text{ch jet}$, as measured by \textsc{Alice} \cite{ALICE:2017nij}. Charged jets are clustered using only charged particles. Interestingly, no modification of the jet mass distribution is found. This is also true for the other jet $\pt$ bins, which are not shown here.

\begin{figure}
\includegraphics[width=0.5\textwidth]{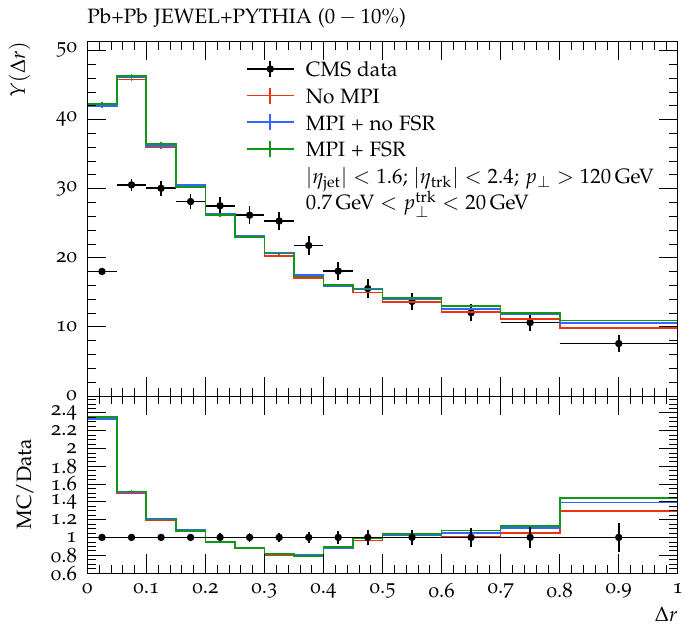}
\caption{Charged-particle track yield $Y$ as a function of the distance $\Delta r = \sqrt{(\Delta \eta)^2 + (\Delta \phi)^2}$ from the jet axis with $R=0.4$ for Pb+Pb collisions, as measured by CMS \cite{CMS:2018zze}.}
\label{fig:pbpb-ydr}
\end{figure}

\begin{figure}
\includegraphics[width=0.5\textwidth]{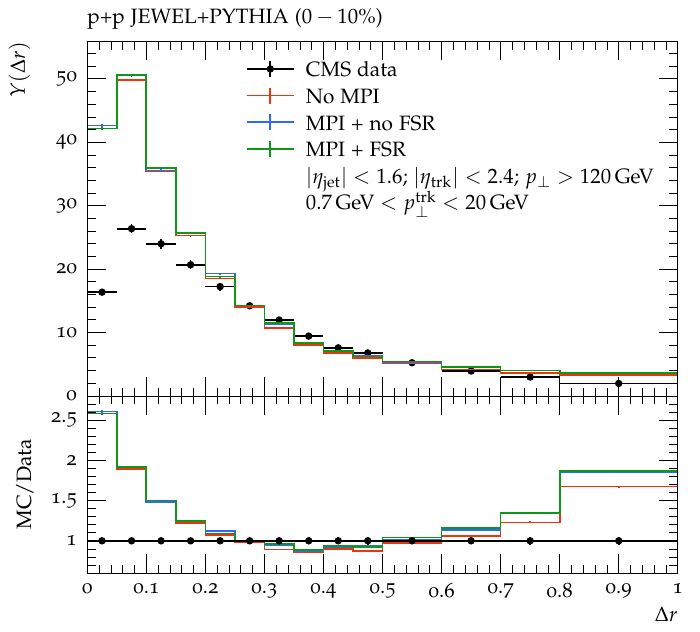}
\caption{Charged-particle track yield $Y$ as a function of the distance $\Delta r = \sqrt{(\Delta \eta)^2 + (\Delta \phi)^2}$ from the jet axis with $R=0.4$ for p+p collisions, as measured by CMS \cite{CMS:2018zze}.}
\label{fig:pp-ydr}
\end{figure}

Jet--hadron correlations can be used to characterise the hadron distribution further away from the jet axis and thus in regions that are less dominated by the jet fragments. Figures \ref{fig:pbpb-ydr} and \ref{fig:pp-ydr} show the charged-particle track yields $Y$ as a function of the distance \\$\Delta r = \sqrt{(\Delta \eta)^2 + (\Delta \phi)^2}$ from the jet axis for Pb+Pb and p+p collisions, respectively. They are compared to measurements by CMS \cite{CMS:2018zze}, where events are selected with at least one jet with $\pt > \unit[80]{GeV}$. As expected, there is a slight enhancement at larger angles from the jet axis due to MPI's, in both the Pb+Pb and p+p spectra.  

\begin{figure}
\includegraphics[width=0.5\textwidth]{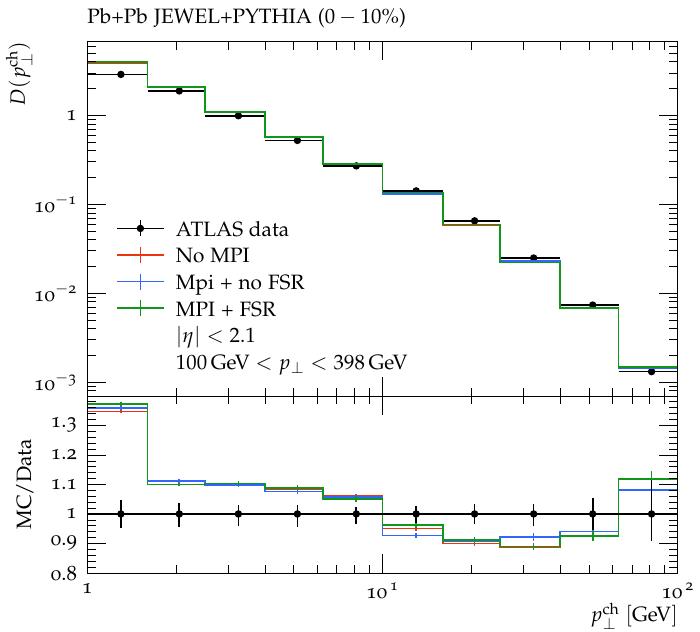}
\caption{Distribution $D$ as a function of charged particle transverse momentum $(\pt^\text{ch})$ with $R=0.4$ for Pb+Pb collisions, as measured by \textsc{Atlas} \cite{ATLAS:2017nre}.}
\label{fig:pbpb-fragfunction}
\end{figure}

\begin{figure}
\includegraphics[width=0.5\textwidth]{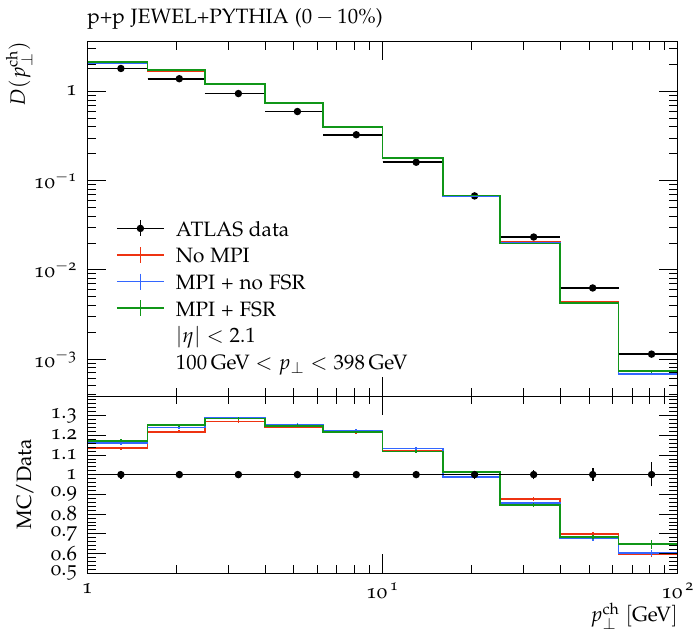}
\caption{Distribution $D$ as a function of charged particle transverse momentum $\pt^\text{ch}$ with $R=0.4$ for p+p collisions, as measured by \textsc{Atlas} \cite{ATLAS:2017nre}.}
\label{fig:pp-fragfunction}
\end{figure}

As a last observable we show the jet fragmentation function, which characterises how the jet momentum is shared among the hadrons that make up the jet. Figures \ref{fig:pbpb-fragfunction} and \ref{fig:pp-fragfunction} show the jet distribution $D$ as a function of the charged-particle transverse momentum $\pt^\text{ch}$ in Pb+Pb and p+p collisions, respectively, as measured by \textsc{Atlas} \cite{ATLAS:2017nre}. It is defined as
%----------
\begin{equation}
D(\pt^\text{ch}) 
:= \frac{1}{N_\text{jet}} \frac{\mathrm{d}N_\text{ch}(\pt^\text{ch})}{\mathrm{d}\pt^\text{ch}},
\label{eq:fragfunction}
\end{equation}
%----------
where $N_\text{ch}$ is the number of charged particles associated with a jet. In p+p collisions, MPI's give rise to a very slight increase at low $\pt$, but no such modification is visible in Pb+Pb collisions. As discussed in the next section, this is probably due to quenching of the MPI partons.
%%----------
%\begin{equation}
%D(z) 
%:= \frac{1}{N_\text{jet}} \frac{\mathrm{d}N_\text{ch}}{\mathrm{d}z},
%\label{eq:fragfunction}
%\end{equation}
%%----------
%where $N_\text{ch}$ is the number of charged particles associated with a jet and 
%%----------
%\begin{equation}
%z
%:= \frac{\pt^\text{ch}}{\pt^\text{jet}} \cos \Delta R
%\end{equation}
%%----------
%is the longitudinal momentum fraction. The distance $\Delta R = \sqrt{(\Delta \eta)^2 + (\Delta \phi)^2}$ is in terms of the relative pseudorapidity $\Delta \eta$ and azimuthal angle $\Delta \phi$ with respect to the jet axis. 

%--------------------------------------------------

\subsection{Di-jet parton-level results}
\label{sec::results:partons}

To gain a better understanding of MPI contributions, we also analyse di-jet events at parton level, where individual partons can be unambiguously assigned to either the hardest scattering or an MPI (which is impossible at hadron level). For the analysis at parton level, jets are reconstructed with a radius $R=0.6$ to make MPI contributions more visible without going to too large radii that cannot be measured experimentally. All parton-level jets have $|\eta| < 3$ and $\pt > \unit[100]{GeV}$.

\begin{figure}
\includegraphics[width=0.5\textwidth]{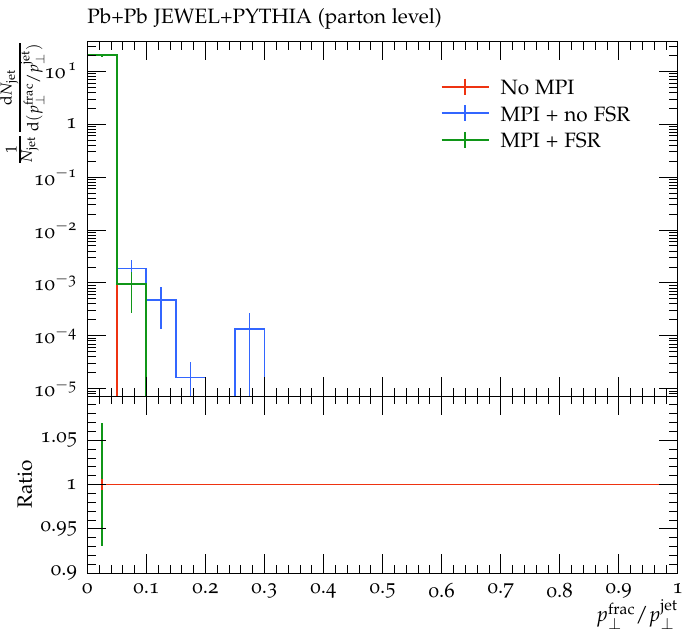}
\caption{Distribution of the fraction $\pt^\text{frac}$ of the total transverse momentum of the jet $\pt^\text{jet}$ that is carried by the MPI partons, shown for $R=0.6$ jets in Pb+Pb collisions.}
\label{fig:pbpb-ptfrac}
\end{figure}

\begin{figure}
\includegraphics[width=0.5\textwidth]{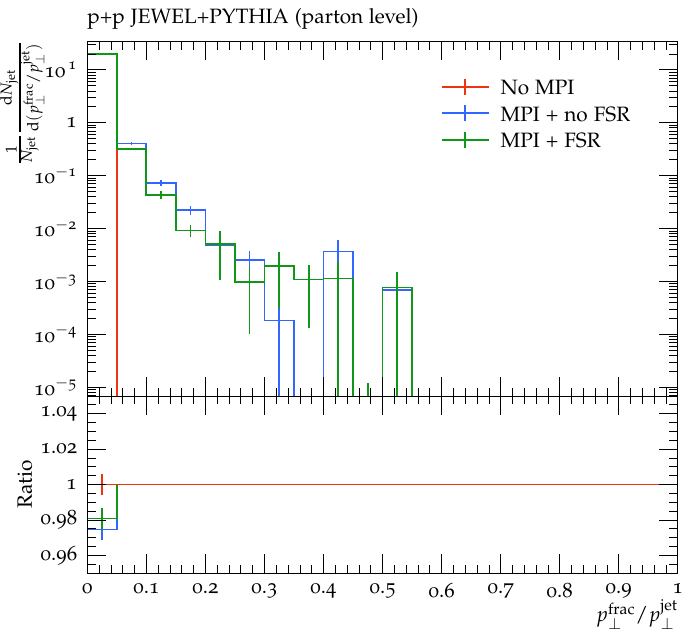}
\caption{Distribution of the fraction $\pt^\text{frac}$ of the total transverse momentum of the jet $\pt^\text{jet}$ that is carried by the MPI partons, shown for $R=0.6$ jets in p+p collisions.}
\label{fig:pp-ptfrac}
\end{figure}

The distributions for the fraction $\pt^\text{frac}$ of the total transverse momentum of the jet $\pt^\text{jet}$ that is carried by the MPI partons in Pb+Pb and p+p collisions is shown in figures \ref{fig:pbpb-ptfrac} and \ref{fig:pp-ptfrac}, respectively. In both cases, the distribution peaks at very small values. This means that the jet $\pt$ is carried almost exclusively by partons coming from the hardest scattering and MPI's do not produce additional jets with $\pt > \unit[100]{GeV}$. Interestingly, the MPI contribution to the jet $\pt$ is significantly larger in p+p than in Pb+Pb. MPI's are, by construction, softer than the hardest scattering and MPI partons thus get quenched more in heavy-ion collisions. Quenching effects distribute the energy of MPI partons broadly in phase space and make it less likely than in p+p collisions that there is enough energy in the form of MPI partons within the jet cone to give a sizable contribution to the jet $\pt$. As seen in figure~\ref{fig:pbpb-ptfrac}, this effect is more pronounced when the MPI partons have final-state parton showers that distribute the energy among more partons and amplify the quenching effect.

\begin{figure}
\includegraphics[width=0.5\textwidth]{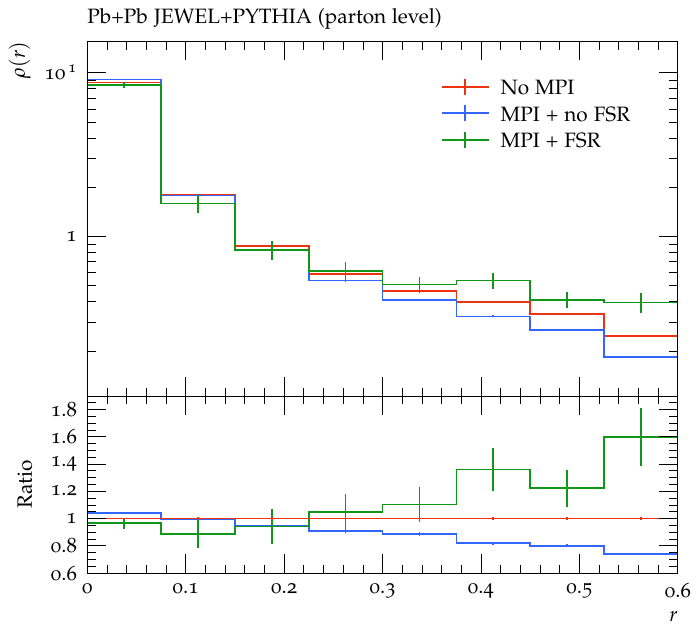} 
\caption{Jet profile $\rho$ as defined in Equation \ref{eq:rho} as a function of the distance $r$ from the jet axis with $R=0.6$, shown for Pb+Pb collisions.}
\label{fig:pbpb-jetprofile}
\end{figure}

\begin{figure}
\includegraphics[width=0.5\textwidth]{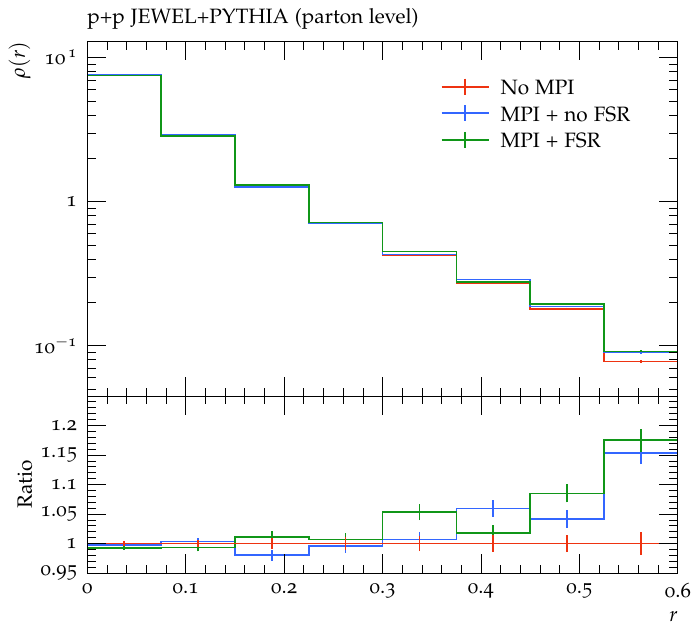} 
\caption{Jet profile $\rho$ as defined in Equation \ref{eq:rho} as a function of the distance $r$ from the jet axis with $R=0.6$, shown for p+p collisions.}
\label{fig:pp-jetprofile}
\end{figure}

Figures \ref{fig:pbpb-jetprofile} and \ref{fig:pp-jetprofile} show the jet profile $\rho (r)$ for Pb+Pb and p+p collisions, respectively. The jet profile is the fraction of the jet's transverse momentum contained in an annulus of size $\delta r$ located at a distance $r$ from the jet axis. It is defined as
%----------
\begin{equation}
\rho (r) := \frac{1}{\pt} 
\sum_{\substack{k \mathrm{ \; with} \\ 
\Delta R_{kJ} \in [r, r + \delta r]}} 
\pt^{(k)},
\label{eq:rho}
\end{equation}
%----------
where $\pt$ and $\pt^{(k)}$ are the transverse momenta of the jet and particle $k$, respectively. The sum is taken over all particles in the event, not only over the jet constituents.
$\Delta R_{kJ} := \sqrt{(\Delta \phi_{kJ})^2 + (\Delta \eta_{kJ})^2}$ is the angular separation between particle $k$ and the jet axis.

In p+p collisions (figure~\ref{fig:pp-jetprofile}) the MPI contributions show up as a small enhancement at large distances from the jet axis. In Pb+Pb collisions, however, the sample with MPI but without FSR off MPI's shows the opposite behaviour and falls below the results without MPI at large $r$. This is an effect of medium response and the corresponding subtraction, because this behaviour is not observed when medium response is turned off. When FSR is included for MPI's, the jet profile increases at large $r$ and at the same time, the statistical uncertainty increases significantly. Again, the effect is not seen without medium response and also not in smaller radius jets with medium response. What is happening here is that with MPI's, FSR off MPI's and related medium response, there are so many partons distributed broadly in the event that it is very rare that individual jets can increase their $\pt$ significantly by incorporating many of these uncorrelated partons. The sample then contains a handful of jets that have a far too large weight for their $\pt$ and that dominate the distributions and inflate the statistical uncertainties. This effect is more pronounced at larger jet radii.

\begin{figure}
\includegraphics[width=0.5\textwidth]{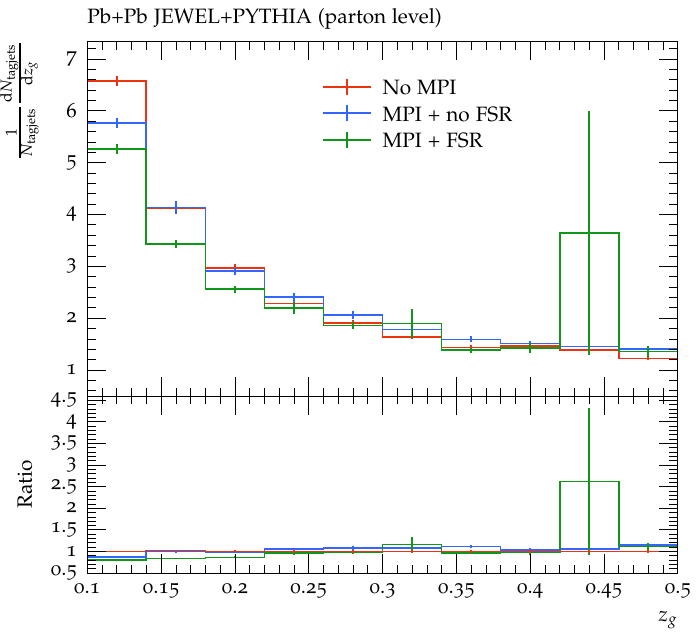}
\caption{Distribution of the SoftDrop variable $z_g$ as defined in Equation \ref{eq:zg}, shown for $R=0.6$ jets Pb+Pb collisions.}
\label{fig:pbpb-zg}
\end{figure}

\begin{figure}
\includegraphics[width=0.5\textwidth]{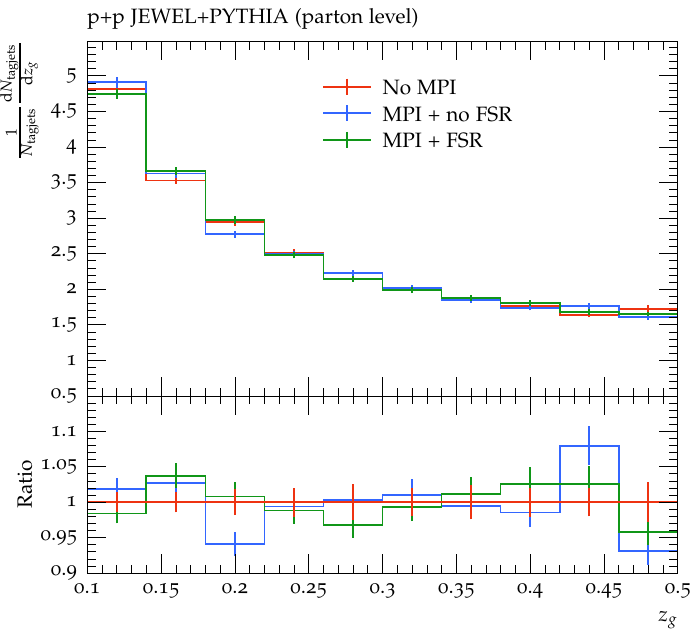}
\caption{Distribution of the SoftDrop variable $z_g$ as defined in Equation \ref{eq:zg}, shown for $R=0.6$ jets in p+p collisions.}
\label{fig:pp-zg}
\end{figure}

In order to gain a more detailed look at the sub-structure of the jets, SoftDrop tagging \cite{Dasgupta:2013ihk,Larkoski:2014wba} is used. The SoftDrop procedure is as follows. First, an anti-$k_\perp$ jet is re-clustered with the Cambridge/Aachen algorithm \cite{Wobisch:1998wt}. In an iterative procedure, the clustering is undone, thereby splitting the jet into two sub-jets. The softer of the two sub-jets is dropped at each step until a configuration is reached that satisfies
%----------
\begin{equation}
z_g := \frac{\min (\pt^{(1)}, \pt^{(2)})}{\pt^{(1)} + \pt^{(2)}}
> z_\mathrm{cut} \left( \frac{\Delta R_{12}}{R} \right)^\beta,
\label{eq:zg}
\end{equation}
%----------
where $\Delta R_{12}$ is the angular separation between the two sub-jets and $\pt^{(i)}$ are their transverse momenta. The $z_g$ distributions for Pb+Pb and p+p collisions are shown in figures \ref{fig:pbpb-zg} and \ref{fig:pp-zg}, respectively. 

\begin{figure}
\includegraphics[width=0.5\textwidth]{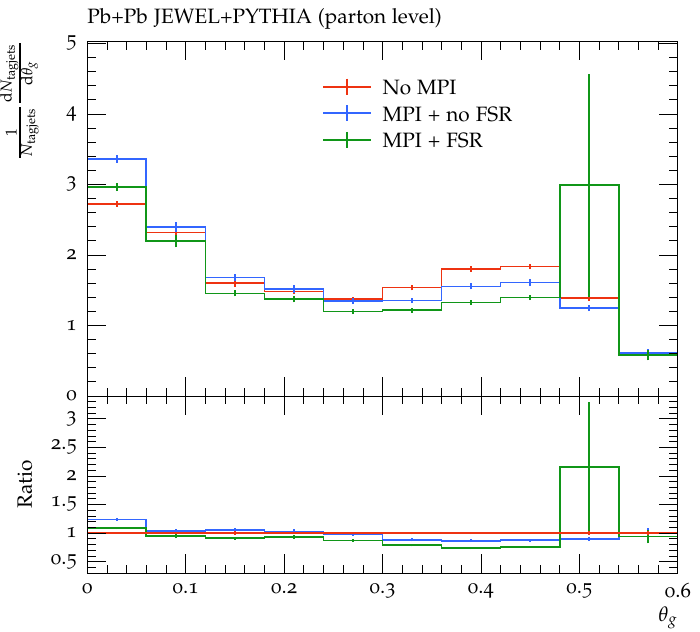}
\caption{Distribution of the SoftDrop opening angle $\theta_g$ between two sub-jets for $R=0.6$ jets in Pb+Pb collisions.}
\label{fig:pbpb-thetag}
\end{figure}

\begin{figure}
\includegraphics[width=0.5\textwidth]{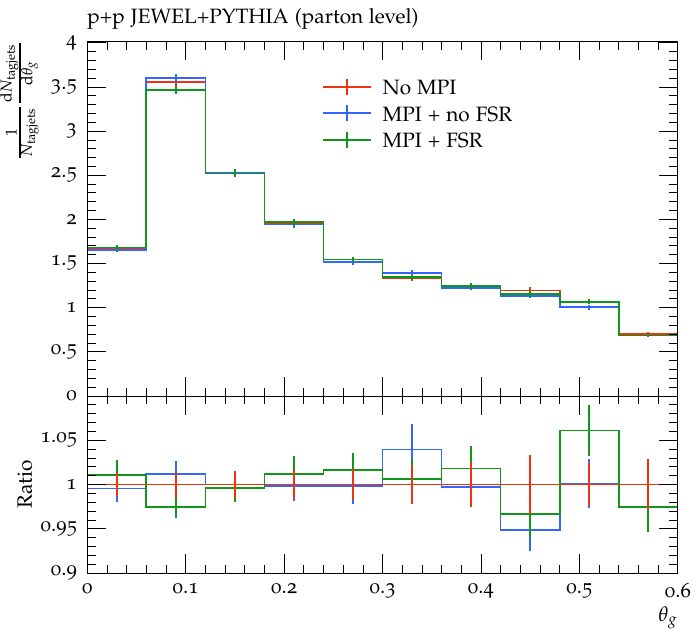}
\caption{Distribution of the SoftDrop opening angle $\theta_g$ between two sub-jets for $R=0.6$ jets in p+p collisions.}
\label{fig:pp-thetag}
\end{figure}

Figures \ref{fig:pbpb-thetag} and \ref{fig:pp-thetag} show the distributions for the opening angle $\theta_g$ between the two sub-jets in the SoftDrop algorithm for Pb+Pb and p+p collisions, respectively. The general observations are similar to the jet profile: in p+p collisions and Pb+Pb collisions without medium response (not shown here), there is no modification of the distributions due to MPI's. In the sample with MPI's but without FSR off MPI's there is a moderate modification that is caused by the interplay of MPI's, medium response and subtraction. The MPI\,+\,FSR sample shows a similar behaviour but has one bin with very large bin value and error bars. This is probably caused by a single jet in the sample that happened to gain a sizable amount of $\pt$ from uncorrelated partons. Because these distributions are normalised, the large value of this one bin pushes the other bins down.

%--------------------------------------------------

\section{Conclusions}

The effect of MPI's on various jet observables in $Z$+jet and di-jet events has been studied here. In many cases, the MPI's do not make a significant change in the distributions. However, in the $Z$+jet case, an enhancement is clearly seen in the angular separation distributions, both in the Pb+Pb and the p+p case. This is in agreement with the results found in \cite{Yang:2021qtl}. Jet--hadron correlations show a small increase at large distances from the jet axis, but no sizable modification is observed in jet $R_\text{AA}$, jet fragmentation distributions or jet mass. 

At the partonic level, it is seen that quenching effects tend to suppress the MPI contribution compared to p+p collisions. Jets with larger radii have erratic behavior due to their large size. This is because they can gain a sizable amount of $\pt$ by sweeping up uncorrelated partons from MPI's and the corresponding parton showers and medium response, and then have a too large weight for their $\pt$. This is very rare but introduces huge fluctuations that hinder the interpretation of the results. There are also indications that the interplay of MPI's, medium response and subtraction introduces artefacts. This was not observed at hadron level and would require further dedicated studies. However, given that the MPI contributions were generally found to be very small, it is questionable whether such a study is worthwhile.

This is the first systematic investigation of MPI's in jet quenching to our knowledge. We have used the rather simple old \textsc{Pythia}\,6 model for the MPI's. There is clearly room for improvements of the modeling, but as a first indication this should be sufficient to show what types of effects one can expect from MPI's. Since the conclusion of this study is that MPI effects in quenched jets are generally negligible, even in jet sub-structure and jet shape observables, investing in a better MPI model hardly seems worthwhile.

\begin{acknowledgements}
This study is part of a project that has received funding from the European Research Council (ERC) under the European Union’s Horizon 2020 research and innovation programme (Grant agreement No. 803183, collectiveQCD).
\end{acknowledgements}

\bibliographystyle{spphys}       % APS-like style for physics
\bibliography{references}   % name your BibTeX data base

\end{document}